%% file: concur.tex
\documentclass[letter]{ieice}
\usepackage[dvips]{graphicx}
\field{}
\title{A  Concurrent Model for Imperative Languages with Improved Atomicity}
\authorlist{
\authorentry{Keehang KWON}{m}{labelA}
\authorentry{Daeseong Kang}{n}{labelB}
}
\affiliate[labelA]{The authors are with Computer Eng., DongA University. email:khkwon@dau.ac.kr}
\affiliate[labelB]{The author is with Electronics Eng., DongA University.}
\received{2003}{1}{1}
\revised{2003}{1}{1}
\finalreceived{2003}{1}{1}

\makeatletter
\long\def\@makemyfntext#1{$^{\rm *}\ $ #1}

\long\def\@myfootnotetext#1{\insert\footins{\footnotesize
    \interlinepenalty\interfootnotelinepenalty 
    \splittopskip\footnotesep
    \splitmaxdepth \dp\strutbox \floatingpenalty \@MM
    \hsize\columnwidth \@parboxrestore
   \edef\@currentlabel{\csname p@footnote\endcsname\@thefnmark}\@makemyfntext
    {\rule{\z@}{\footnotesep}\ignorespaces
      #1\strut}}}

\def\myfootnotetext{\@ifnextchar
     [{\@xfootnotenext}{\xdef\@thefnmark{\thempfn}\@myfootnotetext}}
\makeatother

\input macros

\newcommand{\muprolog}{{C$^{\|}$}}

\newcommand{\kch}{\|}
\newcommand{\sqs}{;}
\newcommand{\blk}{\sharp}

\begin{document}
\maketitle
\begin{summary}
We propose a  new concurrent  model for imperative languages where concurrency occurs 
at a subprogram level.
 This model  introduces a new {\it block sequential} statement of the form $\blk(G_1,\ldots,G_n)$ 
 where each $G_i$ is a statement. This statement tells the machine to execute 
 $G_1,\ldots,G_n$ sequentially and
atomically (\ie, without interleaving). It therefore enhances atomicity and predictability in
 concurrent programming.

 We illustrate our idea
via \muprolog, an extension of the core concurrent C  with the new  block sequential statement.

\end{summary}
\begin{keywords}
concurrency,  imperative languages, block sequential.
\end{keywords}

\section{Introduction}\label{sec:intro}

 Adding concurrency to imperative  programming -- C, its extension \cite{KHP13} and Java -- is an attractive task. 
While concurrent programming may appear to be a simple task, it has proven too difficult to use,
predict or debug.

An analysis shows that this difficulty comes from the fact that 
 interleavings among threads -- which are typically done by OS schedulers --  are quite arbitrary and 
cannot be controlled at all  by  the programmer.

We observe that concurrent programming can be made easier by making interleaving less arbitrary and
more controlled. Inspired by \cite{Jap03,Jap08}, we propose two ideas for this.
First, we propose a concurrent model with its own embedded scheduler.
       Our scheduler works on a high-level: it 
allows the execution to switch from one assigment  statement to another.
This is in contrast to OS schedulers which 
allows the execution to switch from one machine instruction to another.
For example, the assignment statement $c = c+1$ is  $atomic$ in the sense that
  it runs to completion without being interleaved.
This reduces nondeterminism and error known as {\it transient} errors.

Second, semaphores and monitors are typically used as facilities for mutual exclusion.
     We propose a simpler method to solve  mutual exclusion. 
    Toward this end, we propose a new {\it block sequential} statement 
$\blk(G_1,\ldots,G_n)$, 
where each  $G_i$ is a statement. This has the following execution semantics:
execute $G_1,\ldots,G_n$ sequentially and consecutively (\ie, without being interleaved). 
In other words, our interpreter  treats $\blk(G_1,\ldots,G_n)$  atomic. 
This can be easily implemented by designing our interpreter working in two different modes:
the $concurrent$ mode and the traditional $sequential$ mode. Thus, if our interpreter encounters 
$\blk(G_1,\ldots,G_n)$, it switches from the concurrent mode 
to the sequential mode and proceeds just like the traditional
C interpreter. In this way, atomicity of this statement is guaranteed.
We intend to use this construct to each critical region.


This paper focuses on the minimum 
core of C, enhanced with concurrency at a subprogram level. 
This is to present the idea as concisely as possible.
The remainder of this paper is structured as follows. We describe 
 \muprolog, an extension of concurrent C  with a new 
 statement
 in Section 2. In Section \ref{sec:modules}, we
present an example of  \muprolog.
Section~\ref{sec:conc} concludes the paper.

\section{The Language}\label{sec:logic}

The language is   core C 
 with  procedure definitions. It is described
by $G$- and $D$-formulas given by the syntax rules below:
\begin{exmple}
\>$G ::=$ \>   $true \sep A \sep x = E \sep  \sqs(G_1,\ldots,G_n) \sep   $ \\  
\>\>  $\blk(G_1,\ldots,G_n)$ \\
\>$D ::=$ \>  $ A = G\ \sep \all x\ D$\\
\end{exmple}
\noindent
 In the above, 
$A$  represents a head of an atomic procedure definition of the form $p(x_1,\ldots,x_n)$ 
where $x_1,\ldots,x_n$ are parameters.
A $D$-formula  is called a  procedure definition.
In the transition system to be considered, a $G$-formula will function as a thread and
a set of $G$-formulas (\ie, a set of threads) will function as the
main  statement, and a set of $D$-formulas  enhanced with the
machine state (a set of variable-value bindings) will constitute  a program.
Thus, a program is a union of two disjoint sets, \ie, $\{ D_1,\ldots,D_n \} \cup \theta$
where each $D_i$ is a $D$-formula and $\theta$ represents the machine state.
Note that $\theta$ is initially set to an empty set and will be updated dynamically during execution
via the assignment statements. 

 We will  present an interpreter for our language via a proof theory \cite{Khan87,MNPS91,HM94,MN12}.
This is in contrast to other complex approaches to describing an interpreter for
concurrent languages \cite{Alg14,BP09}.
Note that  our interpreter  alternates between 
 the concurrent execution phase 
and the backchaining phase.  
In  the concurrent execution phase (denoted by $ex(\Pscr,\kch(\Gamma,G,\Delta),\Pscr')$) it tries to select and
execute a thread $G$ among a set of threads ($\Gamma,G,\Delta$) with respect to
a program $\Pscr$ and
produce a new program $\Pscr'$
by reducing $G$ 
to simpler forms until $G$ becomes an assignment statement, true  or a procedure call. The rules
 (4)-(11) deal with this phase. Here both $\Gamma$ and $\Delta$ denote a set of $G$-formulas.
If $G$ becomes a procedure call, the interpreter switches to the backchaining mode. This is encoded in the rule (3). 
In the backchaining mode (denoted by $bc(D,\Pscr,A,\Pscr',\Gamma,\Delta)$), the interpreter tries 
to solve a procedure call  $A$ and produce a new  program $\Pscr'$
by first reducing a procedure definition $D$ in a program $\Pscr$ to  its instance
 (via rule (2)) and then backchaining on the resulting 
definition (via rule (1)).
 To be specific, the rule (2) basically deals with argument passing: it eliminates the universal quantifier $x$ in $\all x D$
by picking a value $t$ for
$x$ so that the resulting instantiation, written as $[t/x]D$, matches the procedure call $A$.
 The notation $S$\ seqand\ $R$ denotes the  sequential execution of two tasks. To be precise, it denotes
the following: execute $S$ and execute
$R$ sequentially. It is considered a success if both executions succeed.
Similarly, the notation $S$\ parand\ $R$ denotes the  parallel execution of two tasks. To be precise, it denotes
the following: execute $S$ and execute
$R$  in any order.  Thus, the execution order is not important here. 
It is considered a success if both executions succeed.
The notation $S \leftarrow R$ denotes  reverse implication, \ie, $R \rightarrow S$.

\begin{defn}\label{def:semantics}
Let $\kch(G_1,\ldots,G_n)$ be a sequence of threads to run concurrently and let $\Pscr$ be a program.
Then the notion of   executing $\lb \Pscr,\kch(G_1,\ldots,G_n) \rb$ concurrently and producing a new
program $\Pscr'$-- $ex(\Pscr,\kch(G_1,\ldots,G_n),\Pscr')$ --
 is defined as follows:

\begin{numberedlist}

\item    $bc((A = G_1),\Pscr,A,\Pscr_1,\Gamma,\Delta)\ \leftarrow$  \\
 $ex(\Pscr,\kch(\Gamma,G_1,\Delta),\Pscr_1)$. \% A matching procedure for $A$ is found.

\item    $bc(\all x D,\Pscr,A,\Pscr_1,\Gamma,\Delta)\ \leftarrow$  \\
  $bc([t/x]D,\Pscr, A,\Pscr_1,\Gamma,\Delta)$. \% argument passing

\item    $ex(\Pscr,\kch(\Gamma,A,\Delta),\Pscr_1)\ \leftarrow$    $(D \in \Pscr$ parand $bc(D,\Pscr, A,\Pscr_1,\Gamma,\Delta))$. \% $A$ is a procedure call 

\item  $ex(\Pscr,\kch(true),\Pscr)$. \% True is always a success.

\item  $ex(\Pscr,\kch(),\Pscr)$. \%  Empty threads mean a success.


\item  $ex(\Pscr,\kch(\Gamma,x = E,\Delta),\Pscr_1)\ \leftarrow$ \\
  $eval(\Pscr,E,E')$ 
seqand \% evaluate $E$ to get $E'$ \\
$(ex(\Pscr\uplus \{ \lb x,E' \rb \},\kch(\Gamma,\Delta),\Pscr_1)$

\% If an assignment statement $x = E$ is chosen by our interpreter, 
update $x$ to $E'$ and return to the interpreter. Here, 
$\uplus$ denotes a set union but $\lb x,V\rb$ in $\Pscr$ will be replaced by $\lb x,E' \rb$.

\item  $ex(\Pscr,\kch(\Gamma,\sqs(),\Delta),\Pscr_1)\ \leftarrow$ \\
    $ex(\Pscr,\kch(\Gamma,\Delta),\Pscr_1))$. \% an empty sequential composition is a success.

\item  $ex(\Pscr,\kch(\Gamma,\sqs(G_1,\ldots, G_m),\Delta),\Pscr_2)\ \leftarrow$ \\
  $(ex(\Pscr,\kch(G_1),\Pscr_1)$  seqand \\
  $ex(\Pscr_1,\kch(\Gamma,\sqs(G_2,\ldots,G_m),\Delta),\Pscr_2))$. \% 

\% If  a sequential composition $;(G_1,\ldots, G_m)$ is chosen, execute $G_1$ in sequential mode and then
return to the interpreter with the rest.

\item  $ex(\Pscr,\kch(\Gamma, repeat(G),\Delta),\Pscr_2)\ \leftarrow$ \\
  $(ex(\Pscr,\kch(G),\Pscr_1)$  seqand \\
  $ex(\Pscr_1,\kch(\Gamma,repeat(G),\Delta),\Pscr_2))$.  

\% If  a repeat statement  $repeat(G)$ is chosen, execute $G$ in sequential mode and then
return to the interpreter with the rest plus $repeat(G)$. 

\item  $ex(\Pscr,\kch(\blk()),\Pscr)$. \\
     \% An empty block sequential statement is a success.

\item  $ex(\Pscr,\kch(\Gamma,\blk(G_1,\ldots, G_m),\Delta),\Pscr_3)\ \leftarrow$  \\
   $(ex(\Pscr,\kch(G_1),\Pscr_1)$  seqand \\
$(ex(\Pscr_1,\kch(\blk(G_2,\ldots, G_m)),\Pscr_2)$ seqand \\
  $ex(\Pscr_2,\kch(\Gamma,\Delta),\Pscr_3))$.

\% If  a block sequential statement $\blk(G_1,\ldots, G_m)$ is chosen, execute $G_1$ in sequential mode
and then $\blk(G_2,\ldots, G_m)$ in sequential mode and then
return to the interpreter (which proceeds in concurrent mode) with the remaining. \\

\end{numberedlist}
\end{defn}

\noindent
If $ex(\Pscr,G,\Pscr_1)$ has no derivation, then the interpreter returns  the failure.
Initially it works in the concurrent mode. In the concurrent mode, we assume that
our interpreter chooses a thread
using some predetermined algorithm.
Note that executing $one$ thread $G$ concurrently, denoted by $\kch(G)$,
is identical to executing $G$ sequentially.

\section{Examples }\label{sec:modules}

As a well-known example, we examine the system that allows people to sign up for a mailing list.
An example of this class is provided by the
following code where the procedure below adds a person to a list:

\begin{exmple}
\% Procedure signup \\ \\
signup(person) =  \\ 
 ($N = N+1\ \blk\ list[N] = person$)  \% critical section \\
\end{exmple}
\noindent and the  main program consists of two concurrent threads as follows:
\begin{exmple}
    $\kch(signup(tom),signup(bill))$
\end{exmple}
\noindent In the above, we used a more traditional notation
$(G_1\blk,\ldots,\blk G_n)$ instead of  $\blk(G_1,\ldots,G_n)$.

Although the above signup procedure is in fact a critical section, it
 works correctly because no interleaving -- due to the presence of $\blk$ --is allowed in the critical
section. Note that our code is very concise compared to the traditional ones using semaphores.

\section{Conclusion}\label{sec:conc}

In this paper, we proposed a  simple concurrent  model for imperative languages.
 This model  introduces a block sequential statement  $\blk(G_1,\ldots,G_n)$ 
 where each $G_i$ is a statement. This statement executes  $G_1,\ldots,G_n$ sequentially and
atomically. It therefore  enhances atomicity and predictability.
 
Although we focused on a simple concurrent model for imperative language at  a subprogram level, it
 seems  
possible to apply our ideas
to existing concurrent and parallel computing models \cite{Alg14,BP09}.

\section{Acknowledgements}

This work  was supported by Dong-A University Research Fund.

\bibliographystyle{ieicetr}



\end{document}

%% file: macros.tex
\newenvironment{numberedlist}
{\begin{list}{\makebox[20pt]{\hss(\arabic{itemno})\enspace}}
             {\usecounter{itemno}\labelwidth 20pt}}{\end{list}}

\newcounter{itemno}

\newcounter{itemno1}

\newcounter{itemno2}

\newcounter{exno}

\newcounter{defno}

\newenvironment{defn}{\refstepcounter{defno}\medskip \noindent {\bf
Definition \thedefno.\ }}{\medskip}

\newcommand{\sep}{\;\vert\;}

\newcommand{\oprove}{\vdash\kern-.6em\lower.7ex\hbox{$\scriptstyle O$}\,}

\newcommand{\Pscr}{{\cal P}}

\newcommand{\pderivation}{{\cal P}\kern -.1em\hbox{\rm -derivation}}
\newcommand{\pderivationl}{{\cal P}\kern -.1em\hbox{\em -derivation}}
\newcommand{\pderivable}{{\cal P}\kern -.1em\hbox{\rm -derivable}}
\newcommand{\pderivablel}{{\cal P}\kern -.1em\hbox{\em -derivable}}
\newcommand{\pderivations}{{\cal P}\kern -.1em\hbox{\rm -derivations}}
\newcommand{\pderivability}{{\cal P}\kern -.1em\hbox{\rm -derivability}}

\newcommand{\all}{\forall}

\newcommand{\ie}{{\em i.e.}}

\newsavebox{\lpartfig}
\newsavebox{\rpartfig}

\newenvironment{exmple}{
 \begingroup \begin{tabbing} \hspace{2em}\= \hspace{3em}\= \hspace{3em}\=
\hspace{3em}\= \hspace{3em}\= \hspace{3em}\= \kill}{
 \end{tabbing}\endgroup}

\newcommand{\lb}{\langle}
\newcommand{\rb}{\rangle}

     {\\* \hspace*{\fill} \end{trivlist}}